\documentclass[12pt]{article}
\usepackage{verbatim,color,bbm,epsfig}
\usepackage[usenames,dvipsnames]{xcolor}
\usepackage{fancyhdr}
\usepackage[authoryear, sort]{natbib}
\usepackage{wasysym}
\usepackage{float}
\usepackage{amssymb}

\def\K{{\cal K}}

\def\W{{\cal W}}

\def\boxit#1{\vbox{\hrule\hbox{\vrule\kern6pt  \vbox{\kern6pt#1\kern6pt}\kern6pt\vrule}\hrule}}

\def\sumi{\hbox{$\sum_{i=1}^n$}}

\def\wh{\widehat}

\def\log{\hbox{log}}

\def\bse{\begin{eqnarray*}}
	\def\ese{\end{eqnarray*}}
\def\be{\begin{eqnarray}}
\def\ee{\end{eqnarray}}
\def\bq{\begin{equation}}
\def\eq{\end{equation}}
\def\bse{\begin{eqnarray*}}
	\def\ese{\end{eqnarray*}}
\def\pr{\hbox{pr}}
\def\wh{\widehat}
\def\trans{^{\rm T}}

\def\th{^{th}}
\def\b1e{{\mathbf e}}

\def\bX{{\mathbf X}}

\def\bW{\W}

\def\boldbeta{\boldmath\beta}

\def\boldtheta{\boldmath\theta}

\def\trans{^{\rm T}}

\def\th{^{th}}
\def\b1e{{\mathbf e}}

\def\bW{{\mathbf W}}

\def\bX{{\mathbf X}}

\def\bY{{\mathbf Y}}

\def\bZ{{\mathbf Z}}

\def\bW{\W}
\def\sumb{\hbox{$\sum_{b=1}^{B}$}}

\def\doutb{${\cal D}_{out}^{(b)}$}
\def\dinb{${\cal D}_{in}^{(b)}$}

\renewcommand\footnoterule{\kern-3pt \hrule \textwidth 2in \kern 2.6pt}

\def\boxit#1{\vbox{\hrule\hbox{\vrule\kern6pt \vbox{\kern6pt \textcolor{blue}{#1}\kern6pt}\kern6pt\vrule}\hrule}}

\def\authorfootnote#1{{\let\thefootnote\relax\footnotetext{#1}}}

\begin{document}

\thispagestyle{empty}
\baselineskip=28pt
\begin{center}
{\LARGE{\bf Finite Sample Hypothesis Tests for Stacked Estimating Equations}}

\baselineskip=16pt

\vskip 5mm

Eli. S. Kravitz\\
Department of Statistics, Texas A\&M University, 3143 TAMU, College Station, TX 77843-3143, USA, kravitze@tamu.edu\\
\hskip 5mm \\
Raymond J. Carroll\\
Department of Statistics, Texas A\&M University, 3143 TAMU, College Station, TX 77843-3143, USA and School of Mathematical and Physical Sciences, University of Technology Sydney, Broadway NSW 2007, Australia, carroll@stat.tamu.edu\\
\hskip 5mm \\
David Ruppert\\
School of Operations Research and Information Engineering and Department of Statistics and Data Science, Cornell University, Ithaca NY 14853, USA, dr24@cornell.edu
\end{center}

\vskip 10mm
\begin{center}
	{\Large{\bf Abstract}}
\end{center}
\baselineskip=18pt
Suppose there are two unknown parameters, each parameter is the solution to an estimating equation, and the estimating equation of one parameter depends on the other parameter. The parameters can be jointly estimated by ``stacking" their estimating equations and solving for both parameters simultaneously. Asymptotic confidence intervals are readily available for stacked estimating equations. We introduce a bootstrap-based hypothesis test for stacked estimating equations which does not rely on asymptotic approximations. Test statistics are constructed by splitting the sample in two, estimating the first parameter on a portion of the sample then plugging the result into the second estimating equation to solve for the next parameter using the remaining sample. To reduce simulation variability from a single split, we repeatedly split the sample and take the sample mean of all the estimates. For parametric models, we derive the limiting distribution of sample splitting estimator and show they are equivalent to stacked estimating equations.
\baselineskip=12pt
\par\vfill\noindent
\underline{\bf Key Words}:
asymptotic theory; bootstrap; estimating equations; exact test; M-estimation; 

\par\medskip\noindent
\underline{\bf Short title}: Hypothesis Tests for Stacked Estimating Equations

\clearpage\pagebreak\newpage
\pagenumbering{arabic}
\newlength{\gnat}
\setlength{\gnat}{22pt}
\baselineskip=\gnat

\section{Introduction} \label{sec1}
\subsection{Stacked Estimating Equations}

Suppose ${W}_1, \ldots, {W}_n$ are independent identically distributed random variables.  An M-estimator, $\wh{\theta}$, solves the vector valued equation
\begin{equation}
0 = n^{-1} \sumi \Psi({W}_i; \theta).\label{eq:DR_Mest1}
\end{equation}
M-estimation was introduced by Peter Huber \citep{huber1964robust, huber1967behavior}, and he established the asymptotic properties of these estimators. \cite{liang1986longitudinal} extended M-estimators to the use of longitudinal data under the name \textit{generalized estimating equations}. An overview of M-estimation is given by \cite{stefanski2002calculus}.

We investigate a type of M-estimators called  \textit{stacking estimating equations} \citep[Appendix A.6.6]{carroll2006measurement}. Suppose $\wh{\theta}$ is the solution to (\ref{eq:DR_Mest1})
%the estimating equation
%\bse
%0 = n^{-1} \sumi \Psi({ W}_i, \theta) 
%\ese
and suppose there is another estimating equation which depends on $\wh{\theta}$ to estimate an additional unknown parameter $\beta$,
\begin{equation}
0 = n^{-1} \sumi \K(Y_i, \wh\theta, \beta,).\label{eq:DR_Mest2}
\end{equation}
Then $\beta$ and $\theta$ can be estimated jointly by ``stacking" their respective estimating equations into a single equation,
\begin{equation}
	0  = \sumi \{ \Psi(W_i, \theta)\trans, \K(Y_i, \theta, \beta) \trans\}\trans,\label{eq:DR_Mest3}
\end{equation}
and  finding $(\wh \theta, \wh \beta)$ by solving for $(\theta, \beta)$ ``simultaneously.''  ``Simultaneously'' is in quotes because (\ref{eq:DR_Mest1}) is assumed to have a unique solution, so that  $\wh \theta$ is determined by the first component of (\ref{eq:DR_Mest3}) and then, with $\theta$ fixed at $\wh \theta$, the second component finds $\wh \beta$.
These estimating equations can be extended to the case where $\beta$ does not depend directly on $\theta$ but instead on a completely specified function of $\theta$ and a vector of covariates, $\bX$. We denote this function by $f(\bX; \theta)$. The estimating equation becomes
\bse
	0  = \sumi \Big\lbrack \Psi(W_i, \theta)\trans, \K\{ Y_i , f(\bX_i; \theta),\beta \} \trans \Big\rbrack \trans.
\ese

Standard M-estimator theory can be applied to this stacked estimating equation to determine the asymptotic covariance matrix, $\Sigma$, of the unknown parameters $(\theta, \beta)$. Wald test statistics are constructed from the parameter and variance estimates. The validity of this test statistic depends directly on how well $\Sigma$ has been estimated. Since variance estimates are based on asymptotic theory, they may not be valid in real-world settings where sample sizes are not large enough to justify asymptotic theory or where regularity conditions are not satisfied. Therefore, Wald tests may have unreliable rejection rates and decreased power.

We propose a test statistic that uses sample splitting combined with bootstrapping to generate exact p-values. We calculate $\wh{\theta}$ using a portion of the dataset, and calculate $\wh{\beta}$ using $\wh{\boldtheta}$ and the remaining sample.  This process can be repeated many times, splitting the sample into different subsets each time to get different parameter estimates which are then combined by taking the sample mean. Hypothesis tests are performed with a bootstrap by simulating from the null distribution of the estimates.

In Section \ref{sec:related_work} we give an overview of previous work in sample splitting. Section \ref{sec:motivation} gives a motivating real data example. Section \ref{sec:basics} details the sample splitting algorithm using estimating equations. Section \ref{sec:asymptotics} reviews asymptotic theory from \cite{Kravitz2019b} where asymptotic equivalence between our method and standard methods for stacking estimating equations is established. In Section \ref{sec:small_sample}, we detail our bootstrap test statistic and provide simulations showing our test statistic has correct level and comparable power to the uniformly most powerful  invariant test, the F-test \citep{ neyman1928use, neyman1933ix}.  In Section \ref{sec:data_analysis} we use sample splitting to create a physical behavior index and use it to predict mortality. Technical details are collected in the appendix.

\section{Related Work} \label{sec:related_work}

Sampling splitting has been studied in the high-dimensional statistical literature for  performing hypothesis testing after variable selection. \cite{wasserman2009high} proposed an ad-hoc ``screen-and-clean'' procedure. Their method is as follows: Assume $(X_i, Y_i), \dots, (X_n, Y_n)$ are independent observations from the linear regression model, $Y_i = X_i \trans \beta + \epsilon_i$, where $\epsilon_i \sim N(0, \sigma^2)$. Partition your sample, $\{1, \dots, n\}$, into two subsets, $D_{in}$ and $D_{out}$ such that $D_{in} \cup D_{out} = \{1, \dots, n\} $ and $D_{in} \cap D_{out} = \emptyset$. Here $\emptyset$ denotes the empty set. Using $D_{in}$ perform a variable selection procedure to select a set of active predictors, $\wh S_n$. Using the active predictors, perform ordinary least squares. Discard all variables that do not have a least-squares estimate that is significant at some prespecified value, say the Bonferroni-corrected critical value of a t-distribution. The authors show their method  controls the Type 1 error rate asymptotically. That is, it selects irrelevant variables no more than a nominal $\alpha$ proportion of the time.
	
\cite{dezeure2015high} point out a flaw in the method of \cite{wasserman2009high} which they refer to as the ``p-value lottery'' . The authors note that hypothesis tests from single-sample splitting are sensitive to the data selected into $D_{in}$ and $D_{out}$. The authors demonstrate that, using only a \underline{single} dataset, the p-values assigned by \cite{wasserman2009high} can vary from 0 to 1. The conclusions one draws are dependent on the split performed before analysis and are therefore prone to simulation error.
	
As a solution to the p-value lottery, \cite{meinshausen2009} suggest performing many splits.  \cite{meinshausen2009} show that a p-value for multiple splits can be calculated as follows: Suppose there are $b=1,\dots,B$ splits. Using the training sample, for each split $b$, construct $p_b$, the p-value for testing that $\beta=0$. Define $Q(\gamma)$ to be the minimum of $1.0$ and the empirical $\gamma\th$ quantile of $(p_1,\dots,p_B)$. Choose a minimum such quantile $\gamma_{\min}$, which might be $\gamma_{\min} = 0.05$. Define the aggregated p-value as
\be
	p = \min[1.0,\{1-\log(\gamma_{\min})\} \min_{\gamma \in (\gamma_{\min},1)} \{Q(\gamma)/\gamma\}]. \label{mb01}
\ee
The authors show that (\ref{mb01}) controls the family-wise error rate. 
	
We suggest two improvements. First, \cite{meinshausen2009} only guarantee the family-wise error rate will be less than the nominal error rate. In practice, their procedure is extremely conservative and the actual family-wise error rate will generally be much lower than the nominal value. The test is especially conservative when the number of parameters is small. We ran $100,000$ null simulations and found that the Meinshausen tests had a level of around 0.0001, rejecting the scalar valued null hypothesis just 12 times. In Section \ref{sec:small_sample}, we suggested replacing (\ref{mb01}) with a bootstrap based hypothesis test that has proper level and reasonable power.
	
%We also provide a framework for consistent parameter estimation using sample splitting. Previously sample splitting has been focused entirely to hypothesis testing. Authors pay little attention to variability in the parameter estimates from split to split. This is likely because the authors focus on variable selection or prediction rather than traditional inference. However, proper estimation is a significant concern. Just as \cite{dezeure2015high} point out a p-value lottery, we draw attention to the ``estimation lottery'' that comes about from sample splitting. Parameter estimates vary between splits and must therefore be aggregated in a sensible way to stabilize the final answer. We suggest taking the mean of all the parameter estimates. We show in Section \ref{sec:asymptotics} that the mean of the parameter estimates is a consistent estimation and becomes asymptotically equivalent to stacked estimating equations.

\section{Motivating Example: Physical Activity and Survival} \label{sec:motivation}
	
This work is partly motivated by the creation and analysis of a physical behavior score to predict mortality.  We build the physical behavior score using the NIH-AARP Study of Diet \citep{schatzkin2001design}. Participants self-reported physical behaviors which are then characterized into 8 discrete components. 
A priori, we specify the expected relationship between these physical activities and survival to be consistent with the kinesiology literature. 
Using the training data $\mathcal{D}_{in}$ for a single sample split,  we fit a binary regression model to survival that satisfies these relationships. The 8 components and their marginal models are listed in Table \ref{constraints}. The expected relationships are justified in Section \ref{sec:data_analysis}. We  rescale the fitted predictor values from the logistic model so people with high levels of beneficial activity are assigned a score near 100 and people with low levels of beneficial activity are assigned a score close to 0. We denote the rescaled fitted predictor values as $f(\bX; \wh{\theta})$.
	
We now ask: Is our physical activity score predictive of mortality? We use $f(\bX; \wh{\theta})$ as a predictor in a logistic regression model, along with covariates, $\bZ$. We fit the model
\bse \label{intro_cox}
	\pr(Y_i | \bX_i, \bZ_i) =  H\{ \beta_0  f(\bX_i; \wh{\theta}) + \bZ_i \trans \beta \},
\ese
where $H(\cdot)$ is the logistic distribution function.
	
We want consistent estimates of $\beta_0$ as well as hypothesis tests with proper level and reasonable power. Wald based test statistics are valid for large samples, but may not perform adequately when sample sizes are smaller. We need a method for testing $H_0: \beta_0 = 0$ which is valid for small sample sizes.

\section{Sample Splitting} \label{sec:basics}

We observe data vectors $(Y_i,W_i,\bX_i,\bZ_i)$ that are independent and identically distributed.  Here $Y_i$ and $W_i$ are outcomes, which may be equal to each other, and $\bX_i$ and $\bZ_i$ are covariate vectors. There are two parameters of interest, $\boldtheta$ and $\boldbeta$. The relationship between $W$ and $(\bX,\bZ,\boldtheta)$ is described by a model which has an estimating equation $\Psi(W,\bX,\bZ,\boldtheta)$.  The relationship of $Y$ and $(\bX,\bZ,\boldbeta,\boldtheta)$ is described a model that has by an estimating equation $\K(Y,\bX,\bZ,\boldbeta,\boldtheta)$. We define $\boldtheta$ and $\boldbeta$ as the solutions to $E\{\Psi(W,\bX,\bZ,\boldtheta)\}= 0$ and $E\{\K(Y,\bX,\bZ,\boldbeta,\boldtheta)\}=0$, respectively.

In this case, for a single data set, consistent estimation of $(\boldbeta,\boldtheta)$ can be done by solving the stacked estimating equation
\be
	0 = \sumi \{\Psi\trans(W_i,\boldtheta),\K\trans(\bY_i,\boldbeta,\boldtheta)\}. \label{eq01}
\ee
Asymptotic theory for such estimators is well-known \citep{huber1964robust, huber1967behavior,stefanski2002calculus}. In the example from Section \ref{sec:motivation}, $\K(\cdot)$ does not depend directly on $\theta$ but rather on a function of $\theta$ and $\bX$, denoted with $f(\bX; \theta)$. That is,
\bse
	\K(\bY,\bX,\bZ,\boldbeta,\boldtheta) =  \K(\bY,f(\bX; \boldtheta),\bZ,\boldbeta). \label{eq02}
\ese
Stacked estimating equations and our sample splitting methods apply to this situation, but to simplify notation we do not include $f(\bX; \theta)$ when writing $\K(\cdot)$. The proofs in the Appendix make explicit the dependence of $\K(\cdot)$ on $f(\bX; \theta)$, rather than just $\theta$.

\subsection{A Single Split} \label{sec:single_split}
Let $(\delta_1,...,\delta_n)$ be independent and identically distributed Bernoulli$(\pi)$ random variables. In our applications we set $\pi = 1/2$ so that the splits are approximately equal size. We denote the first partition, ${\cal D}_{in}$, by $\delta = 1$, which estimates $\boldtheta$ by solving
\be
0 = \sumi \delta_i \Psi(\bW_i,\bX_i,\bZ_i,\boldtheta). \label{eq03}
\ee
In the second partition, ${\cal D}_{out}$, denoted by $\delta = 0$, we estimate $\boldbeta$ by solving
\be
0 =  \sumi (1-\delta_i) \K(\bY_i,\bX_i,\bZ_i,\boldbeta,\wh{\boldtheta}). \label{eq04}
\ee

\subsection{Many Splits} \label{sec:many_splits}
In Section \ref{sec:single_split}, $\wh{\beta}$ depends on the particular random sample split. This is an analogue to the variable selection procedure of \cite{wasserman2009high}, who also propose a single split, with variable selection conditioned on the data with $\delta = 1$. \cite{meinshausen2009} and \cite{dezeure2015high} criticize this and call it a ``p-value lottery" as the p-value can vary from 0 to 1 depending on the split. 
Instead, in their context, they suggest using multiple data splits to eliminate simulation variability from using only a single split. 

We define an indicator vector for each sample split, $b$. For $b=1,\dots,B$ and $i=1,\dots,n$, let $(\delta_{1b},\dots,\delta_{nb})_{b = 1}^{B} $ be independent and identically distributed Bernoulli$(\pi)$ random variables. Set $\delta_{ib} = 1$ if the $i\th$ person is selected into the $b^{th}$ training set \dinb. Then solve
\bse
	0 = \sumi \delta_{ib}\Psi(\bW_i,\bX_i,\bZ_i,\boldtheta). \label{eq:psi_b}
\ese
to get an estimate $\wh{\boldtheta}_b$. The subscript denotes the dependence on the parameter estimate of the $b^{th}$ sample split.
	
Now, set $\delta_{ib} = 0$ if the $i\th$ person is selected into the test set, \doutb. We then get an estimate $\wh \boldbeta_b$ by solving
\bse
	0 = \sumi (1-\delta_{ib}) \K(\bY_i,\bX_i,\bZ_i,\boldbeta,\wh{\boldtheta}). \label{eq:kappa_b}
\ese
	
This gives $B$ estimates of $\boldtheta$ and $\beta$. We combine them with the sample mean to get $\wh{\boldtheta} = B^{-1}\sumb \wh{\boldtheta}_b$ and $\wh{\boldbeta} = B^{-1}\sumb \wh{\boldbeta}_b$.

\subsection{Asymptotic Theory} \label{sec:asymptotics}

\cite{Kravitz2019b} develop asymptotic theory for three cases of sample splitting: $B = 1, 1 < B < \infty$, and $B \to \infty$.  They use the theory of ``stacked'' estimating equations \citep[Appendix A.6.6]{carroll2006measurement} and provide asymptotic expansions for each case. Asymptotically valid hypothesis tests can be performed using Wald test statistics and Normal-based confidence intervals.  In a simulation study, they find that coverage probabilties are close to nominal values when $n$ is large, say $n \ge 250$, and $B$ is moderately large, say $B \ge 25$, but coverage probabilities can be well below the nominal values for smaller sample sizes.   Because of this problem, we develop small-sample tests that are exact or nearly exact.

\section{Small Sample Hypothesis Tests} \label{sec:small_sample}

%M-estimator theory can be applied to used for hypothesis testing instead of sample splitting. The validity of this depends on how well the asymptotic covariance matrix, $\Sigma$, has been estimated. Since variance estimates are based on asymptotic theory, they may not be valid in real-world settings where sample sizes are not large enough to justify asymptotic expansions or where regularity conditions are not satisfied. P-values from Wald based tests may be not be correct, leading to inflated rejection rates or decreased power.
%
%While \cite{meinshausen2009} use a test statistic which does not rely on asymptotic normality, it is too conservative for many applications. In the example in Section \ref{sec:motivation}, proper Type I error rates are very important: we do not want to falsely claim our composite score is predictive of mortality when in reality it is not. 

In this section, we propose  a test statistic that uses sample splitting combined with a bootstrap to generate exact p-values, and we  demonstrate through finite sample simulation that it has sufficient power to detect non-null parameters.  

%The asymptotic equivalence between sample splitting and stacking estimating equations established in Section \ref{sec:infinite_split_theory} suggests we can interpret this bootstrap test statistic as a small sample alternative to the Wald Test statistic.

\subsection{A Bootstrap Test Statistic} \label{sec:test_statistic}
The bootstrap algorithm is described below. We give an example of how to implement this algorithm in Supplementary Material \ref{sec:example_algorithm}

\begin{enumerate}
\item \textbf{Sample Splitting:}  Generate $b=1,\dots,B$ sample splits to create training sets, $\mathcal{D}^{(b)}_{in}$, and the test sets, $\mathcal{D}^{(b)}_{out}$

\item \textbf{p-values:} Using the data in $\mathcal{D}^{(b)}_{in}$, solve the first estimating equation to get $\wh\theta_b$. Then solve the second estimation using $D^{(b)}_{out}$ to get the parameter estimate $\wh \beta_{0b}$ and p-value, $p_b$.

\item \textbf{Aggregate:} Take the mean of the parameter estimates: $\wh\beta_{b} := B^{-1} \sum_b \wh\beta_{b}$. Take the mean of the p-values: $p^{H_1} := B^{-1} \sumb p_b$.  

\item \textbf{Null Simulation:}  Simulate $N$ sample from the null distribution of $p^{H_1}$ to get $\widetilde{p}_1, \ldots, \widetilde{p}_N$. Some parameters are not known in the null distribution and must be set to preliminary estimates. This is the case with error variance and non-null parameter estimates.

\item \textbf{New p-value:} Denote $\wh F(\cdot)$ as the empirical distribution of null p-values, calculated from the sample  taken in the last step. Our new p-value is defined as $p^*  = \wh F(p^{H_1}) = $\hbox{$\sum_{i = 1}^N $} $\{ \mathbb{I} (\widetilde{p}_n) < p^{H_1}  \}$, the proportion of null p-values less than the alternative p-value.
\end{enumerate}

\subsection{An Exact Test} \label{sec:exact_test}
	
 We  reject $H_0: \wh \beta_0 = 0$ whenever $p^* < \alpha$ where $p^*$ is the p-value from step 5 above. When the null is true ($\beta_0 = 0$) and there are no unknown nuisance parameters we can draw samples from the exact null distribution of $p^{H_1}$. This will be the case when working with model $Y_i = \beta (X_i \trans \theta) + \epsilon_i$ and the distribution of $\epsilon_i$ is known exactly. Since the null distribution of $Y_i = \epsilon_i$ is known exactly, there are no approximations and the test is exact up to simulation error. This is demonstrated through simulation in Section \ref{sec:power_level}
	
In a more realistic scenario, there will be unknown parameter. We can estimate these quantities then simulate from the approximate null distribution by parametric bootstrap. For linear models we may instead use residual resampling \citep{efron1977efficiency} to sample from the approximate distribution of the error term. In these cases the bootstrap test is not exact. We can, however, get a test that has approximately correct level even though it is not possible to draw samples from the true null distribution of $p^{H_1}$.

\subsection{Level Simulations} \label{sec:power_level}

We want to establish that our bootstrap test statistic is properly leveled.
We assume $E[Y_i|X_i] = \beta_0 (X_i \trans \theta)$ where, as in the real life example in Section \ref{sec:motivation}, we are interested in consistent estimation  hypothesis tests of $\beta_0$. We set $ \vert \vert \theta \vert \vert = 1$, so $\beta_0$ is identified. We express this as two linear models which can be translated into stacked estimating equations. 
\be
Y_i  & = &  \bX_i \trans \theta + \epsilon_i \nonumber \\
Y_i  & = &  \beta_0  (\bX_i \trans \theta) + \epsilon_i,  \label{eq:level_power_model}
\ee
 Under the strong null hypothesis, $Y$ and $\bX$ are independent, e.g., $H_0: \theta = 0 $ or $\beta = 0$. Our null model becomes
\bse
Y_i  =  \epsilon_i \; .
\ese

We simulate $Y$ and $\bX$ under the null hypothesis using a variety of error distributions. We generate independent errors and allow the distribution of the errors to vary. First we assume that the error distribution is known. This allows us to simulate from the true null distribution of $Y|\bX$ and in turn simulate p-values under the null. This is shown in Table \ref{tab:level_known_error}.
		
In most applications, however, the error distribution will not be known, and simulating from the exact null distribution will be impossible. For this case we use residual resampling \cite{efron1994introduction}. A parametric bootstrap is also valid though less robust to misspecified error distributions.  The results are in Table \ref{tab:level_unknown_error}. Our test statistic has approximately proper level for a variety of error distributions and the level remains stable even when the true distribution of the error is unknown.

\subsection{Power Simulations} \label{sec:power_sim}

We now demonstrate that our test statistic has reasonable power. Again we consider the configuration described in (\ref{eq:level_power_model}). Under the assumption of independent Gaussian errors, the F-test is the uniformly most powerful test for the composite null hypothesis: $H_0: \boldtheta  = 0$ or $\boldbeta_1 = 0$. We can compare our test against the F-test as a benchmark. Our bootstrap test cannot have higher power than the F-test, but if it has similar power we know it is a good test.
		
We run $1000$ simulations with a sample size of $n=100$, $\theta = (1, 1, 1) / \sqrt{3}$, and $B= 50$ sample splits, and let $\beta_0$ vary from $-1$ to 1. We assume the  distribution of the error is completely known and use that information in our simulations. The power is estimated as the proportion of times the null is rejected. Table \ref{tab:all_power} shows these results. The maximum difference between the two tests is 0.015.  Our test does occasionally have slightly higher estimated power the F-test as a result of simulation variability.  This indicates the bootstrap test is a reasonable choice of test in terms of power. 

Next we run 1000 simulations using the same configurations without assuming the error distribution is known. We use residual resampling to draw sample from an approximate null distribution. These results can be seen in Table \ref{tab:power_residual}. If the error variance is not assumed to be known, the test still performs well but loses a small amount of power.

\subsection{Choice of $B$} \label{sec:choice_of_b}

\begin{figure}[ht]
\begin{center}
	\includegraphics[scale = 0.75]{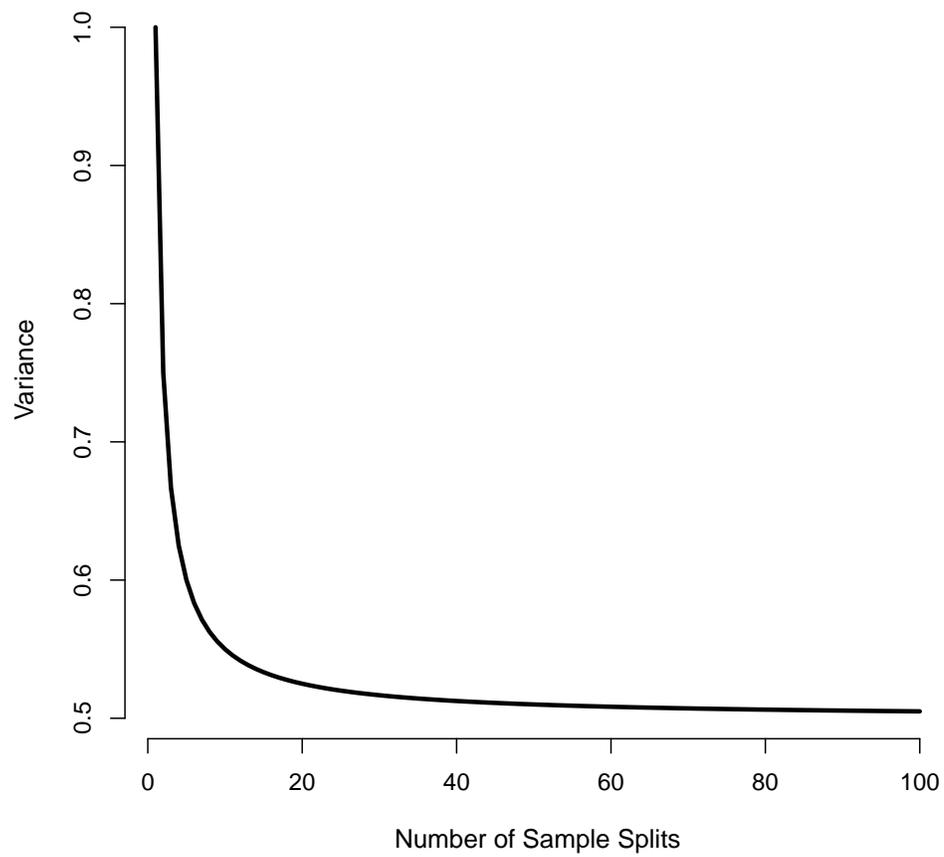}
	\caption{\baselineskip=12pt Variance of a simple sample split estimator of the sample mean. We note that the variance decays very slowly after 50 sample splits.}
	\label{fig:multisplit_var}
\end{center}
\end{figure}

For additional insight, we explore the variance of our test statistic in a simple example of estimating a mean. Suppose
\bse
	Y_i = \mu + \epsilon_i,
\ese
where $\mu$ is the unknown parameter of interest and $\epsilon \sim N(0, \sigma^2)$. In this simple example we only generate \dinb, as this analysis does not require a second estimating equation. In each $\mathcal{D}_{in}^{(b)}$ we use $\bar Y_b$ to estimate $\mu$. Our final estimate of is $\wh \mu = B^{-1} \hbox{$\sum_{b = 1}^B \bar Y_b$}$. Each of $Y_b$ have common correlation $\rho$. In this simple setup  $\rho = 1/2$. The variance of the sample splitting estimate depends on $B$ through,
\be \label{eq:multisplit_vaar}
	{\rm var}(\wh \mu) = \sigma^2/B * \{1 + (B-1)*\rho \}.
\ee
Figure \ref{fig:multisplit_var} plots ${\rm var}(\wh \mu)$ as $B$ increases. The variance decreasing quickly as $B$ increases. When $\sigma = 1$, at our recommended  $B = 50$, ${\rm var}(\wh \mu) = 0.510$ while at $B = \infty$ the variance is only reduced to $0.50$. In most models, $\rho$ cannot be calculated explicitly, but we take from this simulation, along with the power calucaltions, that $B = 50$ is sufficient.

\section{Data Analysis} \label{sec:data_analysis}
 
  Participants in the NIH AARP Study of Diet and Health were ask to complete a questionnaire to measure physical behaviors, medical history, and risk factors for disease. Around a fifth of the total participants (N = 163,106) responded and met criteria for inclusion. Survey responses were translated to time or energy spent in five aerobic activities, two types of sitting activities, and sleep.
 
 We model each of the physical activity components to be consistent with the existing kinesiology literature. The dose-response relationship between aerobic activity and survival has been established as somewhat concave and non-decreasing \citep{arem2015leisure}, with benefits for overall health which level off with increasing activity.  Sedentary time is known to have a negative effect on overall health \citep{grontved2011television, prince2014comparison}.  Sleep is known to be beneficial in reasonable doses, but too much or too little sleep is indicative of poor health	\citep{yin2017relationship}, suggesting a concave, parabolic relationship with overall health. Table \ref{constraints} lists the expected relationships and the marginal models we enforce to describe these relationships.

\subsection{Step 1: Developing the Score}
On each $D_{in}$ we fit the model
\be \label{eq:binary_results}
\hbox{$\pr(W_i = 1 | \bX_i, \bZ_i$) } &=& H \Bigg[ \hbox{$\sum_{j = 1}^{5}$}   \Big\{ d_j - \frac{d_j} {1 + (X_{{\rm aerob}_j} /c_j) ^{b_j}} \Big\} + \theta_{{\rm TV}} X_{{\rm TV}} \nonumber \\
	&+& \theta_{{\rm Sit}} X_{{\rm Sit}} + \theta_{{\rm Sleep}, 1}  X_{{\rm Sleep}} + \theta_{{\rm Sleep}, 2} X^2_{{\rm Sleep}} + \bZ_i \trans \theta\Bigg],
	\ee
where $W_i = 1$ indicates survival until the end of the study, $\bX_i$ is a vector containing the 8 physical activity levels of the $i^{th}$ person, $\bZ_i$ is a vector of covariates including sex, race, education status, and an intercept, and $H(\cdot)$ is the logistic distribution function. 

The first row of Figure \ref{fig:combined_plots}, shows the fitted marginal models for three types of activity: moderate physical activity, sleep, and television sitting. The curves match their intended functional form: moderate physical activity is concave and increasing, sleep is concave, and television sitting is decreasing.

\subsection{Step 2: Rescaling the Score}

The fitted values from (\ref{eq:binary_results}) are the logits of the effect of each type of physical activity on survival. We want to rescale the logits from the physical activity covariates so that their sum is between 0 and 100 with 0 being highest risk and 100 being lowest risk.  While we fit model (\ref{eq:binary_results}) with the additional covariates $\bZ$ to prevent confounding due to demographic information, we do not include the fitted values $\bZ_i \trans \; \wh{\theta}$ when developing the score. 
	
\begin{figure}[ht]
\begin{center}
\includegraphics[width = .9 \textwidth]{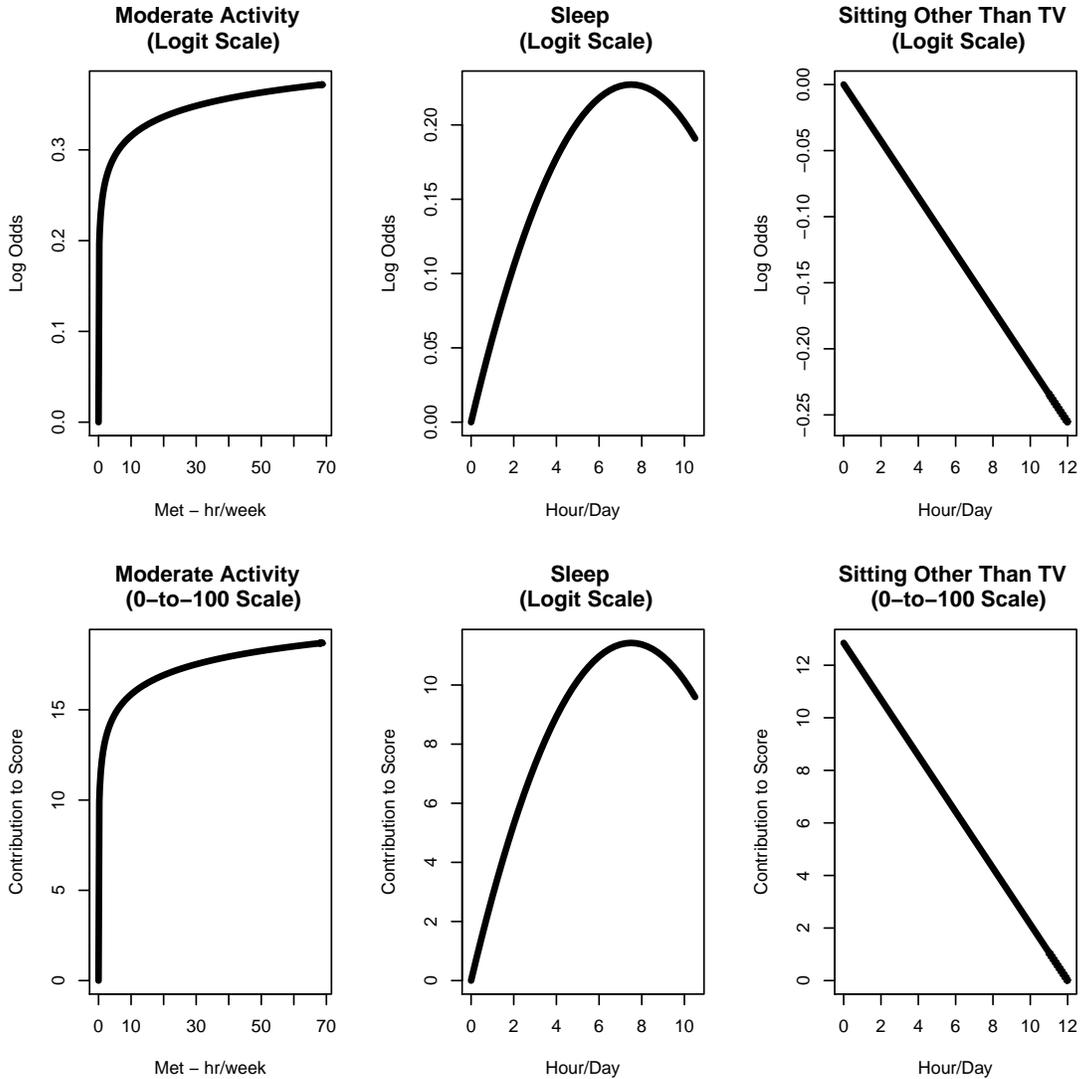}
\caption{\baselineskip=12pt Three of the 8 marginal model plots from binary regression model. The first row shows the marginal models in the original scale and the second rows shows them in a 0-to-100 scale. Moderate activity is modeled to be concave and increase, sleep is modeled to be concave, and television sitting is modeled to be decreasing.  To put the marginal models on a 0-to-100 scale, first make them positive by adding the absolute value of the minimum of each function. Then we rescale the functions so the maximum value of each of the function jointly sum to 100.}
\label{fig:combined_plots}
\end{center}
\end{figure}

To rescale the logits, we first force all the physical activity marginal models to be positive by adding the absolute value of the minimum fitted value, if that value is negative. For example, in the top row of Figure \ref{fig:combined_plots} we see that the marginal model for non-TV sitting is negative for any amount of non-TV sitting greater than 0. The function has a minimum of $-0.25$ at around 12 hours per day. By adding $-0.25$ to the fitted values, we can force this function to always be positive. 
	
Next, we sum the maximum value obtained by each of the, now positive, marginal models and denote this with $T$. We  transform the fitted values with
\be \label{eq:score}
	\frac{T}{100} \bigg[ \hbox{$\sum_{j = 1}^{5}$}   \Big\{ d_j - \frac{d_j} {1 + (X_{{\rm aerob}_j} /c_j) ^{b_j}} \Big\} + \theta_{{\rm TV}} X_{{\rm TV}} \nonumber \\
	+ \theta_{{\rm Sit}} X_{{\rm Sit}} + \theta_{{\rm Sleep}, 1}  X_{{\rm Sleep}} + \theta_{{\rm Sleep}, 2} X^2_{{\rm Sleep}} \bigg]
	\ee
which puts the fitted values from physical activity on a scale from 0 to 100.  We refer to the rescaled marginal models as the contributions to the total score. Three examples of these rescaled marginal models are shown in the bottom row of Figure \ref{fig:combined_plots}. The contribution to the total score is given on the y-axis. Moderate activity, for example, accounts for up to 15 points of the total score of 100. Table \ref{tab:total_score} has an example of the physical activity score using the results from (\ref{eq:score}) on one particular split of the data. Table \ref{tab:total_score} lists the 8 physical behaviors with their relative contribution to a score of 100 and the criteria for receiving a perfect score in each criteria.

\subsection{Step 3: Risk Prediction Based on the Score}

We denote the physical activity score created in the previous section with $f(\bX, \theta)$.  We then estimate the relationship between the physical behavior score and mortality using a Logistic Regression model on \doutb 
\be \label{eq:results_logistic}
\pr(Y_i = 1 | \bX_i, \bZ_i ) &=& H\{ \beta _0 f(\bX_i; \theta ) + \bZ_i \trans \beta \}.
\ee

Our final estimate of $\wh\beta_0$ is $ -0.0026$ with a p-value of 0.003. Our results demonstrate the p-value lottery discussed in Section \ref{sec:related_work}. The p-values from the from splits range from 0.02 to 0.91.  Our parameter estimates vary from $-$0.016 to to 0.012. The relationship between physical activity and survival time is highly significant. Someone with a perfect physical activity score, $f(\bX_i; \theta) = 100$, will be at a 18\% lower risk of all-cause mortality than a person with a physical activity score at the first quintile.

\section{Discussion}\label{sec:discussion_hypothesis_tests}
In this paper we have considered the situation where consistent estimation of one parameter is contingent on estimation of a previous parameter. We suggested splitting the data into two pieces, estimating one parameter on the first subset and using the remaining data to estimate the next parameter.

We developed a theory of sample splitting using estimating equations. 
\cite{Kravitz2019b} provided asymptotic expansions for splitting procedure which can be used to derive asymptotic normality and asymptotically consistent standard errors. 
They found that tests based on asymptotic approximations had p-values below nominal values when the sample was small.
We introduced a bootstrap test statistic which is properly leveled and has nearly optimal power. For sample sizes that are not large enough to justify asymptotic results, we suggest using our sample splitting test statistic.

\section*{Acknowledgments}
Carroll's research was supported by a grant from the National Cancer Institute (U01-CA057030).

\baselineskip=14pt
\bibliographystyle{biomAbhra}
\bibliography{hypothesis_test_July2019}

\newcommand{\Appendix}{\appendix\def\thesection{Appendix~\Alph{section}}\def\thesubsection{\Alph{section}.\arabic{subsection}}}
\section*{Appendix}

\renewcommand{\theequation}{A.\arabic{equation}}
\renewcommand{\thesubsection}{A.\arabic{subsection}}
\setcounter{equation}{0}

\baselineskip=12pt
\clearpage\pagebreak\newpage
\thispagestyle{empty}

\clearpage\pagebreak\newpage
\thispagestyle{empty}
\begin{table}[ht]
\begin{center}
	\begin{tabular}{|l|c|c| }
		\hline\hline
		\textbf{Activity} & \textbf{Expected Relationship} & \textbf{Marginal Model} \\ \hline
		Vigorous Activity & Concave Increasing & 3-parameter Logistic \\*[-0.60em]
		Moderate Activity& Concave Increasing & 3-parameter Logistic  \\*[-0.60em]
		Light Household Activity & Concave Increasing & 3-parameter Logistic  \\*[-0.60em]
		MVPA Household Activity & Concave Increasing& 3-parameter Logistic   \\*[-0.60em]
		Weight Training & Concave Increasing & 3-parameter Logistic  \\*[-0.60em]
		Hours Sitting Other than TV &  Decreasing & Linear \\*[-0.60em]
		Hours of TV Sitting & Decreasing  & Linear\\*[-0.60em]
		Hours of Sleep & Concave  & Quadratic \\
		\hline\hline
	\end{tabular}
	\caption{\baselineskip=12pt Each of the 8 physical activity variables with their respective constraints and marginal model when predicting survival.}
\label{constraints}
\end{center}
\end{table}

\begin{table}[H]
\centering
\begin{tabular}{|c|cccc|}
	\hline \hline
	Distribution\textbackslash{}$B$ & 10     & 25     & 50     & 100    \\ \hline \hline
	Normal            & 0.050 & 0.046  & 0.050 	 & 0.046  \\*[-0.60em]
	t8                & 0.053 & 0.054 & 0.049 & 0.050 \\*[-0.60em]
	t4                & 0.051 & 0.048 & 0.053 & 0.051 \\*[-0.60em]
	Laplace s=1      & 0.058 & 0.049 & 0.053 & 0.049 \\*[-0.60em]
	Laplace s=2      & 0.049 & 0.046 & 0.049 & 0.050 \\*[-0.60em]
	Laplace s=4      & 0.045  & 0.050 & 0.052 & 0.053 \\*[-0.60em]
	N(0,1)N(0,5)p=.1  & 0.045 & 0.058 & 0.050   & 0.051 \\*[-0.60em]
	N(0,1)N(0,10)p=.5 & 0.049 & 0.054  & 0.048 & 0.052 \\ \hline \hline
\end{tabular}
\caption{\baselineskip=12pt Level of the bootstrap test from Section \ref{sec:power_level} for different error distributions and different number of samples splits. The error distribution is completely known}
\label{tab:level_known_error}
\end{table}

\begin{table}[H]
\centering
\begin{tabular}{|c|cccc|}
	\hline \hline
	Distribution\textbackslash{}$B$ & 10 & 25 & 50 & 100 \\ \hline \hline
	Normal & 0.051 & 0.046 & 0.048 & 0.050 \\*[-0.60em]
	t8 & 0.053 & 0.055 & 0.049 & 0.050 \\*[-0.60em]
	t4 & 0.050 & 0.048 & 0.052 & 0.052 \\*[-0.60em]
	Laplace s=1 & 0.056 & 0.053 & 0.049 & 0.047 \\*[-0.60em]
	Laplace s=2 & 0.051 & 0.051 & 0.048 & 0.049 \\*[-0.60em]
	Laplace s=4 & 0.051 & 0.058 & 0.053 & 0.052 \\*[-0.60em]
	N(0,1)N(0,5)p=.1 & 0.045 & 0.059 & 0.054 & 0.053 \\*[-0.60em]
	N(0,1)N(0,10)p=.5 & 0.048 & 0.051 & 0.047 & 0.048 \\ \hline \hline
\end{tabular}
\caption{\baselineskip=12pt Level of the bootstrap test from Section \ref{sec:power_level} using residual resampling for different error distributions and different number of samples splits. The error distribution is not assumed to be known}
\label{tab:level_unknown_error}
\end{table}

\begin{table}[H]
\centering
\begin{tabular}{|c|ccccc|c|}
\hline \hline
$\beta_0$\textbackslash{}$B$ & 10 & 25 & 50 & 100 & 250 & F-Test \\ 
\hline
0 & 0.05 & 0.05 & 0.06 & 0.05 & 0.05 & 0.05 \\*[-0.60em]
0.1 & 0.07 & 0.08 & 0.08 & 0.08 & 0.08 & 0.10 \\*[-0.60em]
0.2 & 0.17 & 0.18 & 0.18 & 0.17 & 0.18 & 0.17 \\*[-0.60em]
0.3 & 0.33 & 0.36 & 0.37 & 0.36 & 0.37 & 0.36 \\*[-0.60em]
0.4 & 0.57 & 0.61 & 0.61 & 0.61 & 0.62 & 0.60 \\*[-0.60em]
0.5 & 0.78 & 0.80 & 0.81 & 0.81 & 0.82 & 0.80 \\*[-0.60em]
0.6 & 0.91 & 0.93 & 0.94 & 0.94 & 0.94 & 0.95 \\*[-0.60em]
0.7 & 0.98 & 0.98 & 0.98 & 0.98 & 0.99 & 0.98 \\*[-0.60em]
0.8 & 0.99 & 1.00 & 1.00 & 1.00 & 1.00 & 1.00 \\*[-0.60em]
0.9 & 1.00 & 1.00 & 1.00 & 1.00 & 1.00 & 1.00 \\*[-0.60em]
1 & 1.00 & 1.00 & 1.00 & 1.00 & 1.00 & 1.00 \\ \hline \hline
\end{tabular}

\caption{\baselineskip=12pt Power of the bootstrap test from Section \ref{sec:power_level} of testing $H_0: \beta_0 = 0$. The unknown parameter $\beta_0$ is varied from 0 to 1. The variance assumed to be known. It is compared to the uniformally most powerful F-test.}
\label{tab:all_power}
\end{table}

\begin{table}[H]
\centering
\begin{tabular}{|c|ccccc|}
\hline \hline
$\beta_0$ \textbackslash{}$B$ & 10 & 25 & 50 & 100 & 250 \\ \hline
0 & 0.05 & 0.05 & 0.05 & 0.05 & 0.04 \\*[-0.60em]
0.1 & 0.07 & 0.08 & 0.08 & 0.08 & 0.07 \\*[-0.60em]
0.2 & 0.17 & 0.18 & 0.18 & 0.16 & 0.18 \\*[-0.60em]
0.3 & 0.33 & 0.36 & 0.36 & 0.36 & 0.37 \\*[-0.60em]
0.4 & 0.57 & 0.61 & 0.60 & 0.60 & 0.61 \\*[-0.60em]
0.5 & 0.78 & 0.80 & 0.81 & 0.81 & 0.81 \\*[-0.60em]
0.6 & 0.91 & 0.93 & 0.94 & 0.93 & 0.94 \\*[-0.60em]
0.7 & 0.98 & 0.98 & 0.98 & 0.98 & 0.99 \\*[-0.60em]
0.8 & 0.99 & 1.00 & 1.00 & 1.00 & 1.00 \\*[-0.60em]
0.9 & 1.00 & 1.00 & 1.00 & 1.00 & 1.00 \\*[-0.60em]
1 & 1.00 & 1.00 & 1.00 & 1.00 & 1.00 \\ \hline \hline
\end{tabular}
\caption{\baselineskip=12pt Power of the bootstrap test from Section \ref{sec:power_level} using residual resampling. All tests are of  $H_0: \beta_0 = 0$. The unknown parameter $\beta_0$ is varied from 0 to 1. The power is very close to when the errors are assumed known, though occasionally lower.}
\label{tab:power_residual}
\end{table}

\begin{table}[ht]
\begin{center}
    \begin{tabular}{|ccc|}
    \hline\hline
    Component                & Contribution to Total & Criteria for Maximum                           \\ \hline  
    Vigorous Activity        & 10                    & \textgreater{}20 MET-hrs/wk                     \\*[-.60em]
    Moderate Activity        & 30                    & \textgreater{}50 MET-hrs/wk \\*[-.60em]
    Light Household Activity & 3                     & \textgreater{}3 MET-hrs/wk                       \\*[-.60em]
    MVPA Household Activity  & 25                    & \textgreater{}20 MET-hrs/wk                      \\*[-.60em]
    Weight Training          & 2                     & \textgreater{} 2 MET-hrs/wk                       \\*[-.60em]
    Sitting Other than TV    & 6                     & \textless{} 3.5 hours                             \\*[-.60em]
    Hours of TV Sitting      & 14                    & \textless{} 2  hours                              \\*[-.60em]
    Hours of Sleep           & 10                     & 7.5 hours \\ \hline                                  
    Total                    & 100                   & \\  \hline\hline                       
    \end{tabular}
    	\caption{\baselineskip=12pt Example physical activity score developed using half of the data. We fit the binary regression model from Section \ref{sec:motivation} and rescale the fitted values of the physical behaviors to be between 0 and 100. The middle column gives the proportion of the total score of 100 that each component can contribute. The third column gives the criteria for receiving the maximum score for each component. MET = metabolic equivalent.}
\label{tab:total_score}
    \end{center}
    \end{table}	

\clearpage\pagebreak\newpage
\thispagestyle{empty}

\clearpage\pagebreak\newpage
\pagestyle{fancy}
\fancyhf{}
\rhead{\bfseries\thepage}
\lhead{\bfseries NOT FOR PUBLICATION SUPPLEMENTARY MATERIAL}
\begin{center}
{\LARGE{\bf Supplementary Material to\\ {\it Finite Sample Hypothesis Tests for Stacked Estimating Equations}}}
\end{center}

\vskip1cm
\begin{center}
Eli. S. Kravitz\\
Department of Statistics, Texas A\&M University, 3143 TAMU, College Station, TX 77843-3143, USA, kravitze@tamu.edu\\
\hskip 5mm \\
Raymond J. Carroll\\
Department of Statistics, Texas A\&M University, 3143 TAMU, College Station, TX 77843-3143, USA and School of Mathematical and Physical Sciences, University of Technology Sydney, Broadway NSW 2007, Australia, carroll@stat.tamu.edu\\
\hskip 5mm \\
David Ruppert\\
School of Operations Research and Information Engineering and Department of Statistics and Data Science, Cornell University, Ithaca NY 14853, USA, dr24@cornell.edu
\end{center}

\setcounter{equation}{0}
\setcounter{page}{1}
\setcounter{table}{1}
\setcounter{section}{0}
\renewcommand{\theequation}{S.\arabic{equation}}
\renewcommand{\thesection}{S.\arabic{section}}
\renewcommand{\thesubsection}{S.\arabic{section}.\arabic{subsection}}
\renewcommand{\thepage}{S.\arabic{page}}
\renewcommand{\thetable}{S.\arabic{table}}
\baselineskip=17pt

\section{Example of Testing Algorithm from Section \ref{sec:test_statistic}} \label{sec:example_algorithm}

\subsection{Variance Known} \label{sec:example_var_known}

Consider the simple model $Y_i = \beta_0 \bX_i \trans \theta + \epsilon_i$, where $\Vert \theta \Vert = 1$ for identifiability and $\epsilon_i \sim N(0,1)$. Suppose you have $i = 1, \ldots n$ observations.

\begin{enumerate}
\item \textbf{Sample Splitting:}  Generate $b=1,\dots,B$ sample splits:  the training sets, $\mathcal{D}^{(b)}_{in}$ , and the test set, $\mathcal{D}^{(b)}_{out}$

\item \textbf{Alternative Hypothesis:} Using the data in $\mathcal{D}^{(b)}_{in}$, fit the model $Y_i = \bX_i \trans \theta + \epsilon_i$ to get $\wh \theta_b$ Scale $\wh \theta_b$ so $\Vert \wh \theta \Vert = 1$  Then using $D_{out}^{(b)}$, use $\bX_i \trans \wh \theta_b$ as a predictor in the model $Y_i = \beta_0 * \bX_i \trans \wh \theta + \epsilon_i$. This gives us $\wh \beta_{0b}$

\item \textbf{Aggregate:} Take the mean of the p-values: $p^{H_1} := B^{-1} \sumb p_b$. Take the mean of the parameter estimates: $\wh\beta_{0b} := B^{-1} \sum_b \wh\beta_{0b}$. 

\item \textbf{Null Simulation:}  The null hypothesis is $\beta_0 = 0$ Repeat the following procedure $N$ times to get $N$ samples from the null distribution of $p^{H_1}$. Denote these as $\widetilde{p}_1 \ldots \widetilde{p}_N$. 
\begin{enumerate}
\item Simulate Response: Under the null $Y_i \sim N(0,1)$. Generate simulated response: $Y^{*}_{i} \sim N(0,1)$ for $i = 1, \ldots n$
\item Null Sample Splitting: Define $\mathcal{D} = \{X, Y^{*}\}$. Split the sample $B$ times to make $\mathcal{D}_{in}^{(b)}$ and $D_{out}^{(b)}$ and get $\wh \beta_0$ and $p^{H1}$ as before. \textbf{Note:} Since $Y^{*}$ and $X$ are independent $\beta_0$ should be close to 0. For a fixed $b$, $p_b \sim U[0,1]$. The mean of $p_1, \ldots p_b$ will not be uniform though, it is a more complicated distribution. It is \textbf{the null distribution of the mean of the p-values}, e.g. $p^{H_1}$.
\end{enumerate}

\item \textbf{A New Test Statistic:} Denote $\wh F(\cdot)$ as the empirical distribution of null p-values simulated in 4(a)-4(b). Our new p-value is defined as $p^*  = \wh F(p^{H_1}) = $\hbox{$\sum_{i = 1}^N $} $\{ \mathbb{I} (\widetilde{p}_n) < p^{H_1}  \}$, the proportion of null p-values less than the alternative p-value.
\end{enumerate}

\subsection{Unknown Variance} \label{sec:example_var_unknown}
We consider the same model as in \ref{sec:example_var_known}. That is, $Y_i = \beta_0 \bX_i \trans \theta + \epsilon_i$, where $\Vert \theta \Vert = 1$ for identifiability and $\epsilon_i \sim N(0,1)$. There are $i = 1, \ldots, n$ samples.

We use bootstrapping residuals instead of simulating $Y^*$ from $N(0,1)$. The process replaces 4(a) and 4(b) with:
\begin{itemize}
\item Get residuals: Fit $Y = X_i \trans \theta + \epsilon_i$ to get residuals, $\wh e_i$. Center the residuals and scale with $\sqrt{n/(n-p-1)}$.
\item Simulate Y: Draw $\wh e_i$ with replacement. Make simulated response $Y^* = \wh e_i$
\item Null Sample Splitting: Define $\mathcal{D} = \{\bX, Y^{*}\}$. Split the sample $B$ times to make $\mathcal{D}_{in}^{(b)}$ and $D_{out}^{(b)}$ and get $\wh \beta_0$ and $p^{H1}$ as before.
\end{itemize}

\end{document}